# Sub-barrier cavitation regime in liquid helium

**Mikhail Pekker and Mikhail N. Shneider[1*],**

[1]*Department of Mechanical and Aerospace Engineering, Princeton University, Princeton, NJ USA*
[*]m.n.shneider@gmail.com

**Abstract**

In this paper, on the basis of the model Schrödinger equation, we consider the tunneling mechanism of cavitation in liquid helium and obtain threshold values of negative pressure as a function of temperature for $^3$He and $^4$He. The results of calculating the surface tension coefficients for flat and curved interfaces, obtained in the approximation of the Lenard-Jones interaction potential, are presented. It is shown that the temperature dependence of the critical pressure at which cavitation begins is stepwise in nature. The obtained critical pressure values are in satisfactory agreement with the experimental data.

## Introduction

The classical description of the nucleation process, that is, the formation of growing bubbles in a liquid in a negative pressure field, is confronted with an insurmountable contradiction at low temperatures. Indeed, from the classical point of view, the number of bubbles of a critical radius generated per unit volume per unit time is proportional to $e^{-W_{cr}/k_B T}$, where $W_{cr}$ is the minimum activation energy required to create a growing bubble, $k_B$ is Boltzmann constant, and $T$ is temperature of liquid. Since $W_{cr}$ does not depend on the temperature of the fluid, as the temperature decreases, a greater negative pressure is required to form bubbles of critical size. However, from experiments [1] it follows that, starting with a certain fixed temperature, the nucleation rate (determined by the critical negative pressure at which cavitation starts) in liquid helium ceases to depend on temperature.

A similar occurrence takes place in chemistry. In classical chemistry, the rate of any chemical reaction $C$ exponentially depends on the ratio of the activation energy $I_{act}$ and the temperature of the reagents $T$: $C \propto e^{-I_{act}/k_B T}$ (the Arrhenius law). It would seem that with such Arrhenius dependence, the reaction rate should decrease with the decreasing temperature of the reagents until its complete cessation. In experiments, indeed with a decrease in temperature, an exponential decrease in the rate of chemical reactions is observed; however, starting from a certain temperature, the reaction rate ceases to fall and remains constant [2, 3].

Both these facts are explained on the basis of quantum mechanics. Chemical reactions at low temperatures, when the Arrhenius law does not work, become possible due to tunneling, as Hund was first to note in 1927 [4]. Subsequently, the effect of sub-barrier tunneling in "cold chemistry" was considered in many articles, both theoretical and experimental (see reviews [2, 3]). The possibility of tunnel nucleation (creation of growing bubbles) in liquid helium was first considered by Lifshits and Kagan [5] and further developed in many articles, see for example [6–10]. In the original work of Lifshits and Kagan [5] and subsequent works [8, 10], liquid helium was considered in the framework of a continuous medium, a constant coefficient of surface tension (independent of pore size and stretching negative pressure), and Bohr's approach in calculating the energy of zero-point oscillations, $E_0 = \hbar\omega_0$, where $\omega_0$ is the oscillation frequency of a variable-mass particle in a potential well (see Part II).

In this paper, by analogy with the stationary Schrödinger equation for a point particle, a wave equation is proposed that describes the tunneling mechanism of cavitation in liquid helium. The eigenvalues and wave functions for various values of the negative pressure in $^4$He and $^3$He are found. It is shown that the temperature dependence of the critical pressure at which cavitation begins is stepwise in nature and, taking into account the dependence of the surface tension coefficient on the radius of the bubble and the extensibility of the liquid in the field of negative pressure, is in satisfactory agreement with the experimental data.

The method for calculating the surface fluid coefficient is given in Appendixes 1–3.

## I. Statement of the problem

Below, as in [5, 8, 10], we will consider the liquid to be continuous and the cavitation bubbles as voids with an infinitely thin boundary, in which the added mass is concentrated, the surface tension coefficient depending on the bubble radius and negative pressure (Appendixes 1,2).

As in [5], consider the Rayleigh bubble. The kinetic energy associated with the movement of the added mass is [11, 7, 12]:

$$E_{kin} = 2\pi R_b^3 \rho \left(\frac{dR_b}{dt}\right)^2, \quad (1)$$

where $R_b$ is the bubble radius and $\rho$ is the density of liquid.

With the assumption that the interaction between the molecules is determined by the Lennard-Jones potential, follow Appendixes 1–3, the potential energy of the cavitation pores is:

$$W_b = 8\pi\sigma_{|P_-|} \int_{d_m}^{R_b} r\xi_b(r)\,dr - \frac{4\pi}{3}|P_-|\left(R_b^3 - d_m^3\right), \quad (2)$$

where $|P_-|$ is the absolute value of negative pressure, and $\sigma_{|P_-|}$ is the coefficient of the surface tension of liquid in the field of negative pressure $P_-$ (see Fig. A4(b)–(d), Appendix 3). In (2), we took into account that the bubble radius $R_b$ cannot be smaller than the effective diameter of the molecule $d_m$. Function $\xi_b$ is presented in Fig. A3, Appendix 2.

For a constant coefficient of surface tension $\sigma_0$ (independent of pore size and negative pressure and neglecting effective diameter of the molecule ($d_m = 0$)), formula (2) is reduced to the form [11]:

$$W_{0,b} = 4\pi\sigma_0 R_b^2 - \frac{4\pi}{3}|P_-|R_b^3\,. \quad (3)$$

In accordance with [5], we formally write the Lagrange function for a pore as:

$$L\left(R_b, \dot{R}_b\right) = E_{kin} - W_b = 2\pi R_b^3 \rho_{|P_-|} \dot{R}_b^2 - 8\pi\sigma_{|P_-|} \int_{d_m}^{R_b} r\xi_b(r)\,dr + \frac{4\pi}{3}|P_-|\left(R_b^3 - d_m^3\right). \quad (4)$$

Dependences of the density of the stretched fluid $\rho_{|P_-|}$ on negative pressure for $^3$He and $^4$He are shown on Fig. A4(a)–(c), Appendix 3.

Considering that in our thin layer approximation, the velocity of the added mass $\sim \partial R_b/\partial t \equiv \dot{R}_b$, from (4) we find the canonical momentum $p_b$:

$$p_b = -\frac{\partial L}{\partial \dot{R}_b} = -2\pi\rho_{|P_-|}R_b^3\dot{R}_b, \tag{5}$$

and the Hamiltonian corresponding to the Lagrangian (4):

$$H(R_b, P_b) = \frac{p_b^2}{2M} + W_b = \frac{4\pi\rho_{|P_-|}R_b^3}{2}\dot{R}_b^2 + 8\pi\sigma_{|P_-|}\int_{d_m}^{R_b} r\xi_b(r)\,dr - \frac{4\pi}{3}|P_-|\left(R_b^3 - d_m^3\right). \tag{6}$$

At a constant surface tension coefficient $\sigma_0$ (not depending on $R_b$ and radius $P_-$) and a constant density of liquid $\rho_0$ (not depending on $|P_-|$), the Hamiltonian (6) coincides with the Hamiltonian introduced in [5]:

$$H_0(R_b, P_b) = \frac{p_b^2}{2M} + W_{0,b} = \frac{4\pi\rho_0 R_b^3}{2}\dot{R}_b^2 + 4\pi\sigma_0 R_b^2 - \frac{4\pi}{3}|P_-|R_b^3. \tag{7}$$

In the original work of Lifshits and Kagan [5] and subsequent works [8, 10], equation (7) was considered in the framework of Bohr's approach in calculating the energy of zero-point oscillations:

$$T_0 = \frac{2\pi}{\omega_0} = \frac{2\pi}{E_0/\hbar} = 2\int_{R_1}^{R_2} \frac{dR_b}{\left(\left(E_0 - 4\pi\sigma_0 R_b^2 + \frac{4\pi}{3}|P_-|R_b^3\right)\frac{2}{4\pi\rho_0 R_b^3}\right)^{1/2}} \tag{8}$$

The radii $R_1$ and $R_2$ in (8) correspond to the turning points of a particle of variable mass $M = 4\pi\rho_0 R_b^3$ in the potential $U = 4\pi\sigma_0 R_b^2 - \frac{4\pi}{3}|P_-|R_b^3$. The approach formulated by Lifshits and Kagan made it possible to estimate the role of tunneling in the nucleation of cavitation bubbles.

Below we abandoned the semi-classical description, which allowed us to find the critical pressure at which cavitation begins in liquid helium and to trace the transition from the Arrhenius nucleation law to the tunnel one. Our results are consistent with experiments [1].

The Hamiltonian (6) can be regarded as the Hamiltonian of a particle with a mass $M$ depending on its coordinate $R_b$ ($M = 4\pi\rho_{|P_-|}R_b^3$), located in a given potential $W_b$. By analogy with quantum mechanics, when the Hamiltonian (6) for a point particle corresponds to the stationary Schrödinger equation, we can assume that equation (6) corresponds to the equation:

$$\frac{1}{R_b^3}\frac{d}{dR_b}R_b^3\frac{d\Psi}{dR_b} + \left(\frac{2\mathrm{M}}{\hbar^2}(E - W_b)\right)\Psi =$$

$$\frac{1}{R_b^3}\frac{d}{dR_b}R_b^3\frac{d\Psi}{dR_b} + \left(\frac{8\pi\rho_{|P_-|}R_b^3}{\hbar^2}\left(E - 8\pi\sigma_{|P_-|}\int_{d_m}^{R_b} r\xi_b(r)\,dr + \frac{4\pi}{3}|P_-|\left(R_b^3 - d_m^3\right)\right)\right)\Psi = 0, \tag{9}$$

where $\Psi$ is the wave function. Equation (9) differs from the usual Schrödinger equation in that the particle is considered to be point-wise in the Schrödinger equation, and the square of the modulus of the wave function describes the probability of being in a given point in space. Whereas in our case, described by equation (9), "particles" correspond to a hollow sphere with an infinitely thin shell with a mass $M = 4\pi\rho_{|P_-|}R_b^3$, depending on its radius, and the square of the modulus of the wave function describes the probability that the shell will have a radius $R_b$. Obviously, if the mass $M$ is constant, equation (9) goes into the usual Schrödinger equation.

Since in the Lennard-Jones model the function $\xi_b$ is universal, that is, it depends only on $R_b/d_m$ (the graph of the function $\xi_b$ is shown in Appendix Fig.A3), it is convenient to introduce the variable $x = R_b/d_m$. Writing the wave function in the form $\Psi = \chi/R_b$, we get:

$$\frac{d^2\chi}{dx^2} + \alpha x^3 \left( E' - \frac{\sigma_{|P_-|}}{\sigma_0} \int_1^x \xi F(\xi)\, d\xi + \beta(x^3 - 1) \right) \chi = 0, \tag{10}$$

where $\alpha = \frac{64\pi^2 d_m^7 \rho_{|P_-|}\sigma_0}{\hbar^2}$, $\beta = \frac{d_m}{6\sigma_0}|P_-|$, $E' = \frac{E}{8\pi\sigma_0 d_m^2}$.

The wave equation corresponding to the Hamiltonian (7) takes the form:

$$\frac{d^2\chi}{dx^2} + \alpha_0 x^3 \left( E' - \frac{x^2}{2} + \beta x^3 \right) \chi = 0, \tag{11}$$

where $\alpha_0 = \frac{64\pi^2 d_m^7 \rho_0 \sigma_0}{\hbar^2}$ and $\rho_0$ is the density of liquid helium at zero negative pressure.

Thus, as for the conventional Schrödinger equation, our task is to find the spectrum of the eigenvalues $E'$and the corresponding eigenfunctions for equation (10) or (11) at a given fixed value of negative pressure $P_-$.

## II. Eigenvalues and eigenfunctions

Equations (10) and (11), like the Schrödinger equation, have eigenvalues that determine the possible energy levels of a particle/thin shell in a potential well.

Tables 1 and 2 show the eigenvalues of energy in degrees Kelvin for equations (10) and (11) for $^4$He and Tables 3 and 4 for $^3$He. Zero mode corresponds, as for the Schrödinger equation, to the ground state, Mode 1 to the first excited state, 2 to the second excited state, and so on. It follows from Tables 1 and 2 that when calculating model (10), the eigenmodes in He4 disappear at lower values of $|P_-|$ than when calculated by model (11). This is indirect evidence in favor of model (10) since the experimentation [1] cavitation starts at $|P_-|{\sim}0.8 - 1$ MPa. The same applies to the calculation of He$^3$. Note that each value of negative pressure $|P_-|$ corresponds to its final set of quantum states; an increase in negative pressure leads to a decrease in the number of states.

**Table 1.** $^4$He. Eigenvalues of energy in degrees Kelvin corresponding to equation (10).

| He4 (10) $\lvert P_-\rvert$ [MPA] | Mode 0 E[K] | Mode 1 E[K] | Mode 2 E[K] | Mode 3 E[K] | Mode 4 E[K] | Mode 5 E[K] |
|---|---|---|---|---|---|---|
| 0.630 | 0.697 | | | | | |
| 0.605 | 0.921 | 2.609 | | | | |
| 0.590 | 1.038 | 2.999 | 4.213 | | | |
| 0.575 | 1.147 | 3.351 | 4.814 | 5.861 | | |
| 0.565 | 1.216 | 3.569 | 5.175 | 6.387 | 7.284 | |
| 0.555 | 1.283 | 3.779 | 5.514 | 6.865 | 7.935 | 8.751 |
| 0 | 3.597 | 10.920 | 16.430 | 21.342 | 25.623 | 29.681 |

**Table 2.** $^4$He. Eigenvalues of energy in degrees Kelvin corresponding to equation (11).

| He4 (11) $\|P_{-}\|$ [MPA] | Mode 0 E[K] | Mode 1 E[K] | Mode 2 E[K] | Mode 3 E[K] | Mode 4 E[K] | Mode 5 E[K] |
|---|---|---|---|---|---|---|
| 2.00 | 10.12 | | | | | |
| 1.75 | 10.71 | 18.92 | | | | |
| 1.53 | 11.18 | 20.7 | 25.43 | | | |
| 1.42 | 11.40 | 21.51 | 26.93 | 30.35 | | |
| 1.33 | 11.59 | 22.14 | 28.04 | 32.11 | 34.92 | |
| 1.26 | 11.71 | 22.58 | 28.88 | 33.33 | 36.66 | 39.07 |
| 0 | 14.00 | 29.74 | 40.59 | 49.59 | 57.52 | 64.66 |

**Table 3.** $^3$He. Eigenvalues of energy in degrees Kelvin corresponding to equation (10).

| He3 (10) $\|P_{-}\|$ [MPA] | Mode 0 E[K] | Mode 1 E[K] | Mode 2 E[K] | Mode 3 E[K] | Mode 4 E[K] | Mode 5 E[K] |
|---|---|---|---|---|---|---|
| 0.246 | 0.6158 | | | | | |
| 0.233 | 0.7958 | 2.092 | | | | |
| 0.222 | 0.9258 | 2.536 | 3.5155 | | | |
| 0.216 | 0.9911 | 2.7491 | 3.8848 | 4.6789 | | |
| 0.210 | 1.0537 | 2.9489 | 4.2173 | 5.1693 | 5.8814 | |
| 0.206 | 1.0969 | 3.0889 | 4.4661 | 5.4947 | 6.3225 | 6.9602 |
| 0 | 2.492 | 7.368 | 11.07 | 14.28 | 17.2 | 19.91 |

**Table 4.** $^3$He. Eigenvalues of energy in degrees Kelvin corresponding to equation (11).

| He3 (11) $\|P_{-}\|$ [MPA] | Mode 0 E[K] | Mode 1 E[K] | Mode 2 E[K] | Mode 3 E[K] | Mode 4 E[K] | Mode 5 E[K] |
|---|---|---|---|---|---|---|
| 0.68 | 6.04 | | | | | |
| 0.55 | 6.61 | 11.83 | | | | |
| 0.49 | 6.86 | 12.73 | 15.68 | | | |
| 0.456 | 7.00 | 13.19 | 16.56 | 18.00 | | |
| 0.426 | 7.10 | 13.58 | 17.25 | 19.78 | 21.56 | |
| 0.406 | 7.17 | 13.85 | 17.70 | 20.45 | 22.50 | 24.00 |
| 0 | 8.59 | 18.21 | 24.85 | 30.36 | 35.19 | 39.58 |

Fig. 1 shows the potential $U_b = \frac{W_b}{k_B} = \frac{\alpha}{k_B}\left(\frac{\sigma_{|P_-|}}{\sigma_0}\int_1^x \xi F(\xi)\, d\xi - \beta(x^3 - 1)\right)$ in degrees Kelvin corresponding to equation (10) and the ground state wave function for $|P_-| = 0.6055$ MPa. This energy corresponds to the negative pressure at which cavitation begins in $^4$He at $T \to 0$ (see Part III). At a helium temperature less than 0.0697 K, cavitation is determined by mode 0. The horizontal lines show energies $E$ in degrees Kelvin (equation (9) of mode 0 corresponding to the zero state in (b) and of mode 1 corresponding to the first excited state). The vertical lines in Figs. (b) and (c) correspond to the values of $x$ at which $U_b = E$. At $x > x_1$, and the wave function decays. At $x > x_2$ the pore radius begins to grow since the surface tension cannot compensate for the tensile forces associated with negative pressure.

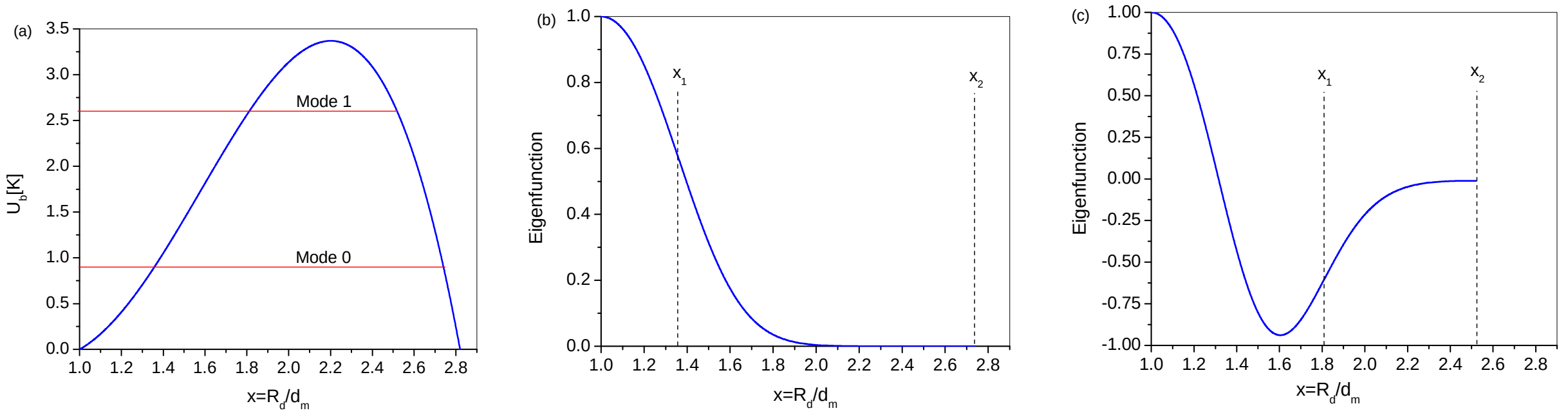

**Figure 1.** (a) – Dependence of the potential $U_b$ on the coordinate $x = R_b/d_m$ for $^4$He; $|P_-| = 0.6055$ MPa. The horizontal lines show energies $E$ in degrees Kelvin (equation (10) of mode 0 corresponding to the zero state in (b) and of mode 1 corresponding to the first excited state). (b)–(d) are the eigenfunctions of equation (10). The gaps $(x_1, x_2)$ corresponds to the attenuation of the wave function.

Fig. 2 shows the potential $U_b$ and the eigenfunctions corresponding to $|P_-| = 0.6$ MPa at helium temperature $T = 0.095$ K; cavitation is determined by the first and second modes, and mode 0 does not contribute to cavitation (see next section).

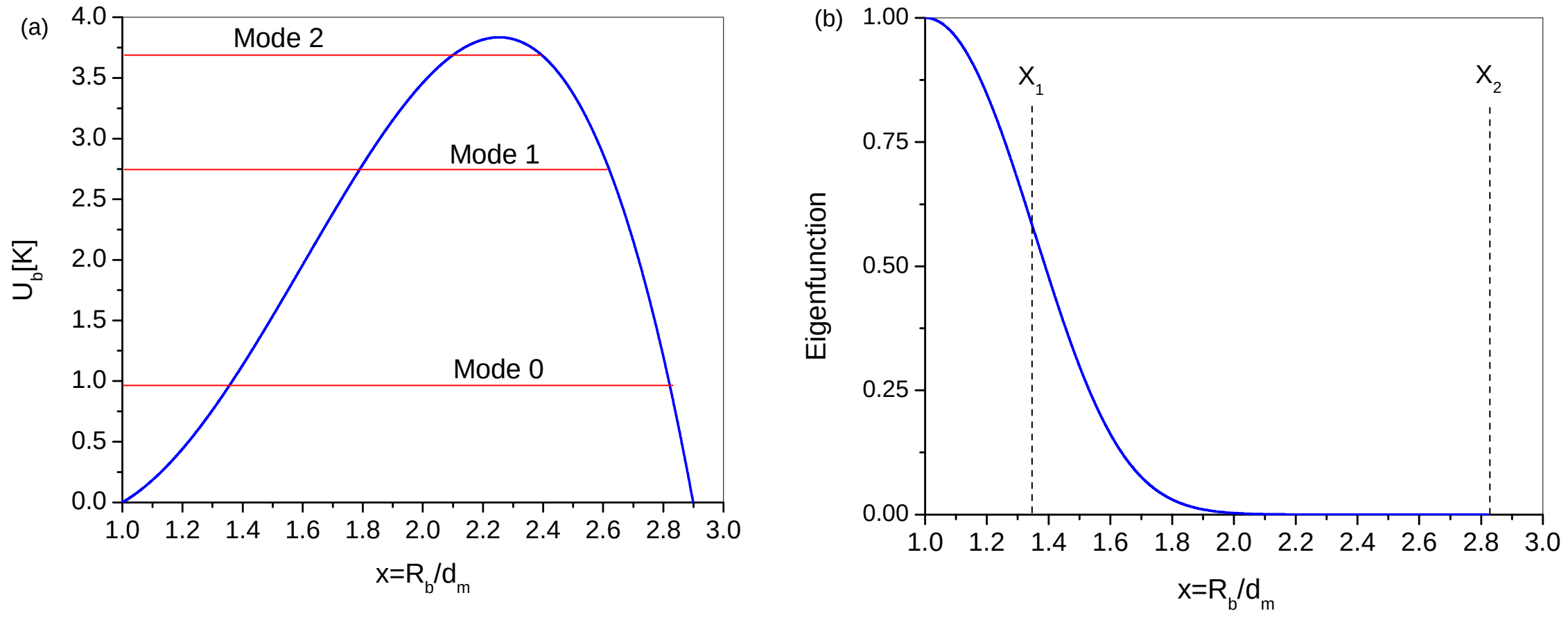

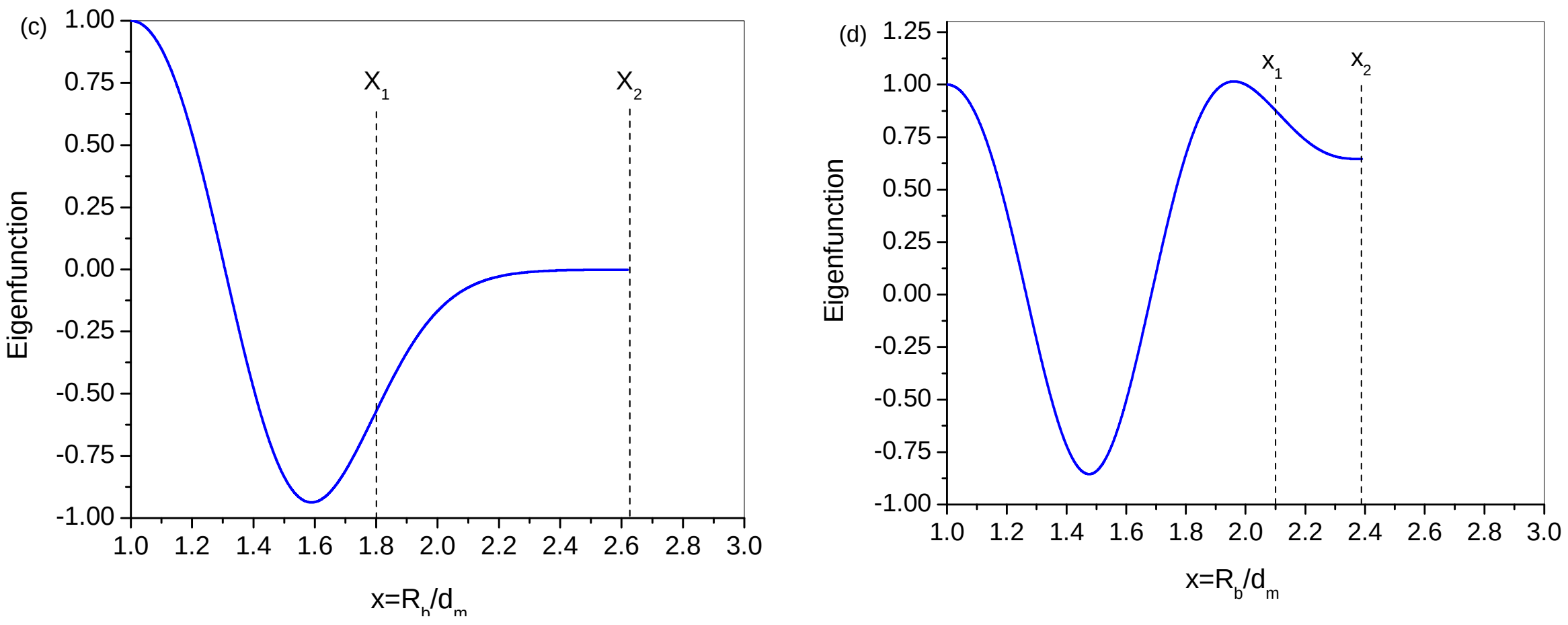

**Figure 2.** (a) – Dependence of the potential $U_b$ on the coordinate $x = R_b/d_m$ for $^4$He; $|P_{-}| = 0.6$ MPa. The horizontal lines show the eigenmodes. (b)–(d) are the eigenfunctions of equation (10). The vertical lines in (b)–(d) correspond to those in Fig. 1.

## III. Nucleation probability

In the classical thermodynamically equilibrium ensemble of micropores, the probability to detect a micropore with energy $E$ is determined by the Boltzmann distribution $F(E) \propto \exp\left(-\frac{E}{T}\right)$ ($E$ in K degree). In this case, if the height of the barrier $U_0$ is much higher than the temperature $T$, then the probability of a particle to leave a potential well (to overcome the barrier) is exponentially small. This is the reason for the above-mentioned sharp decrease in the rate of chemical reactions with a decrease in temperature with the classical Arrhenius temperature dependence (Fig. 3).

Let us turn to the quantum-mechanical consideration. Again consider the "particle" in a potential well. In this well, a particle can have only discrete energy values. Tables 1–4 show these energies obtained by solving equations (10) and (11) for various values of $|P_{-}|$. When $|P_{-}| = 0$, $U_b$ is always a positive value ($\beta = 0$), and the particle cannot be outside the potential well. Tables 1–4 show the set of eigenvalues at $|P_{-}| > 0$, for which a particle can be both inside a potential well and outside it (Fig. 3b). Moreover, if inside a potential well its movement is limited by walls, then outside it is not limited; it can freely move along the $x$ axis. In our case, a cavitation micropore (hollow sphere with an added mass) can unlimitedly increase in size. The difference between the quantum and classical cases is that a particle in a well has a nonzero probability of being outside the well at $x_2$ (Figs. 1b–c, 2b–d) without changing the energy – in other words, to make a tunnel transition.

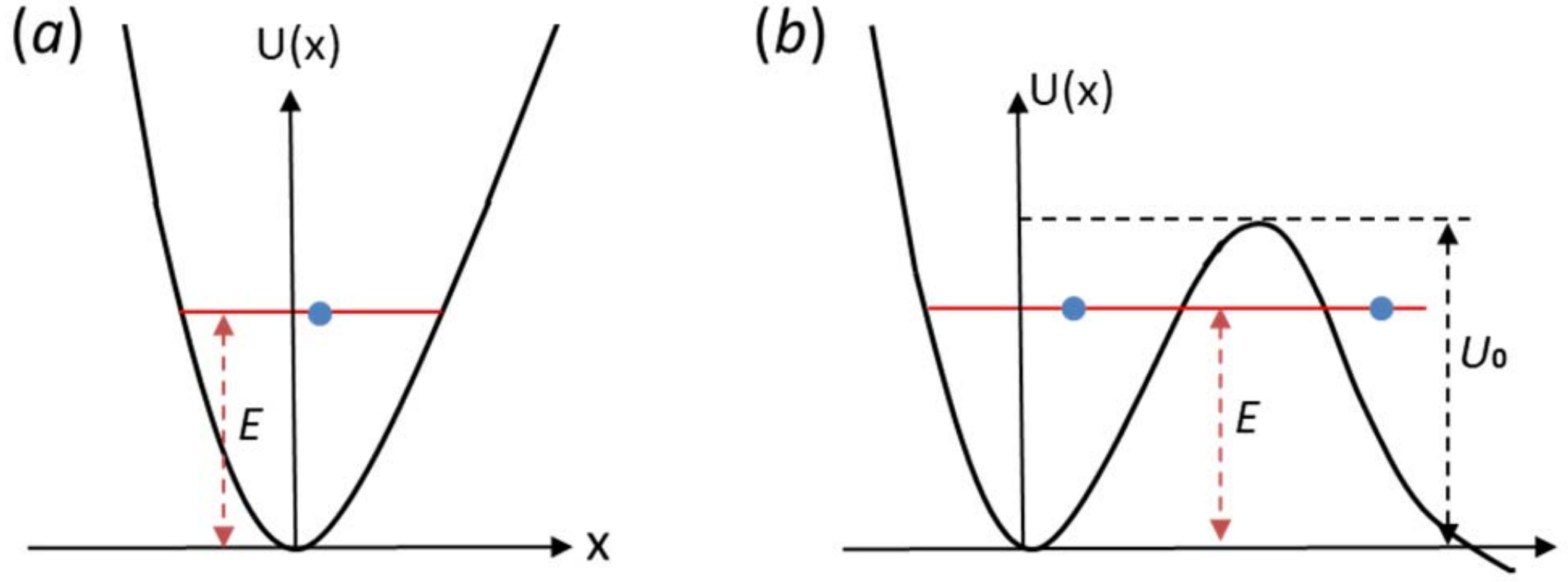

**Figure 3.** A particle in a potential well. (a) – The potential well is infinitely deep. A particle can only be in a potential well. (b) – Potential well with a potential barrier. The particle can be both in the well and outside it. $U_0$ is the height of the potential barrier; $E$ is the energy of the particle.

In accordance with [5], we define the tunneling probability (rate of pore formation of a critical radius) as:

$$\Gamma = k_B \sum_k \frac{E_k}{\hbar} \frac{|\Psi_k(R_{2,k})|^2}{\frac{4\pi}{3}\int_{d_m}^{R_{2,k}} |\Psi_k|^2 r^2 dr} e^{-(E_k-E_0)/T} = k_B \sum_k \frac{E_k}{\frac{4\pi}{3} R_{2,k}^3 \hbar} \frac{|\chi_k(R_{2,k})|^2}{\frac{1}{R_{2,k}}\int_{d_m}^{R_{2,k}} |\chi_k|^2 dr} e^{-(E_k-E_0)/T}, \quad (12)$$

where $k$ is the mode number, $R_{2,k}$ is the pore radius at which its free expansion begins, $E_k$ is the energy of the $k^{\text{th}}$ mode in degrees Kelvin, and $T$ is the temperature of liquid helium. The factor $e^{-(E_k-E_0)/T}$ in (12) indicates the population of the $k$ state. In (12), we took into account that the bubble radius $R_b$ cannot be smaller than the effective diameter of the molecule $d_m$.

In the semi-classical approximation for the case of a particle with constant mass, (12) acquires the form:

$$\frac{|\chi_k(R_{2,k})|^2}{\frac{1}{R_{2,k}}\int_{d_m}^{R_{2,k}} |\chi_k|^2 dr} \approx exp\left(-\frac{2S_k}{\hbar}\right), \quad (13)$$

where $S_k = \int_{R_{1,k}}^{R_{2,k}} (2m(U - E_k)k_B) dr$, and $R_{1,k}$ and $R_{2,k}$ are coordinates of the beginning and end of the potential barrier $U$ for a particle with energy $E_k$. In Fig. 1 (b) - (c) and Fig. 2 (b) - (d) the coordinates$R_{1,k}$, $R_{2,k}$ in dimensionless variables correspond to the points $x_1$ and $x_2$, respectively. (Recall that here $U, E_k$ are in absolute temperature units). Substituting (13) in (12) and introducing the volume of pores $V_{b,k} = \frac{4\pi}{3} R_{2,k}^3$, we obtain Γ in the semi-classical approximation derived by Lifshits and Kagan:

$$\Gamma = \sum_k \frac{k_B E_k}{\hbar} \frac{1}{V_{b,k}} e^{-\frac{2S_k}{\hbar} - (E_k - E_0)/T}. \quad (14)$$

Let us find the temperature limit at which all energy levels, except for the $0^{\text{th}}$ level, can be neglected. В единицах $\mu\text{m}^{-3}\text{ns}^{-1}$ Γ имеет вид:

$$\Gamma = \frac{k_B}{\frac{4\pi}{3} d_m^3 \hbar} 10^{-27} \sum_k \frac{E_k}{x_{2,k}^2} \frac{|\chi_k(x_{2,k})|^2}{\int_1^{x_{2,k}} |\chi_k|^2 dx} e^{-\frac{(E_k - E_0)}{T}} \, [\mu\text{m}^{-3}\text{ns}^{-1}] \, . \qquad (15)$$

Here $x_{2,k} = R_{2,k}/d_m$. Defining the beginning of cavitation as in [12–15] by condition: $\Gamma > \Gamma_{cr} = 1\ \mu\text{m}^{-3}\text{ns}^{-1}$, we find that the temperature of liquid helium $T$, at which the contribution of all modes to (15), except of the $0^{\text{th}}$ mode, can be neglected. With an increase in negative pressure, the height of the potential barrier decreases (compare Figs. 1a and 2a), and the number of possible modes decreases. As $T \to 0$, all terms in the sum in (15), except for the first one, corresponding to the zero mode, can be neglected. The critical negative pressures at which cavitation initiates for $^3$He and $^4$He are $|P_{cr}| =$ 0.2315 and 0.6054 MPa, respectively. Fig. 4 shows the calculated dependences of the critical pressure on the temperature of liquid helium. The steps in the graph correspond to the appearance of new modes with a decrease in negative pressure, which can be interpreted as a discrete change in the pore size corresponding to the possible discrete allowable eigenvalues of the solution of equation (10). For example, for $^4$He and $^3$He the critical pressure does not change within the temperature ranges 0–0.0697 K and 0–0.0581 K, correspondingly.

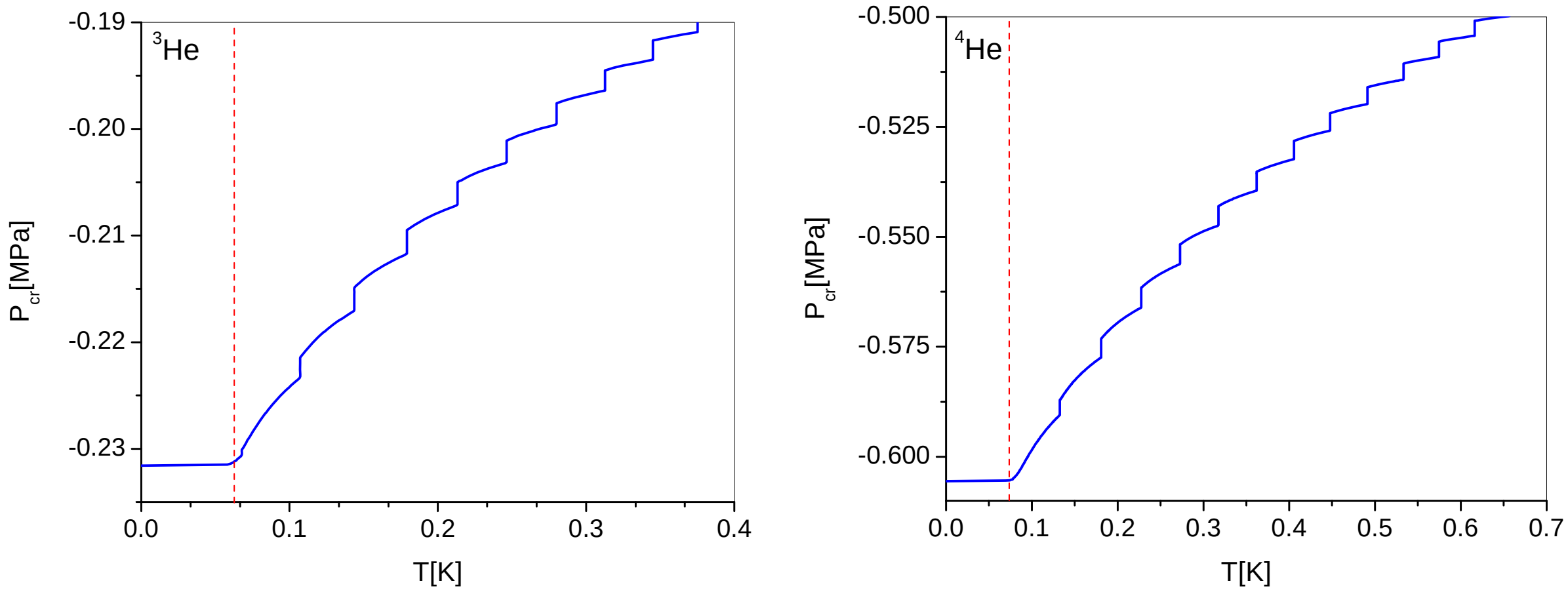

**Figure 4.** Calculated dependences of the critical pressure on temperature for $^3$He and $^4$He.

Fig. 5 shows the dependences of the critical pressure on the temperature of liquid helium calculated on the basis of (11). As can be seen, the critical pressure differs by more than a factor of two from the calculation based on the solution of equation (10). For $T \to 0$, a critical pressure $P_{cr} = -0.578$ MPa for $^3$He (92% below the lower limit) and $P_{cr}$ =-1.79 MPa for $^4$He (79% below the lower limit).

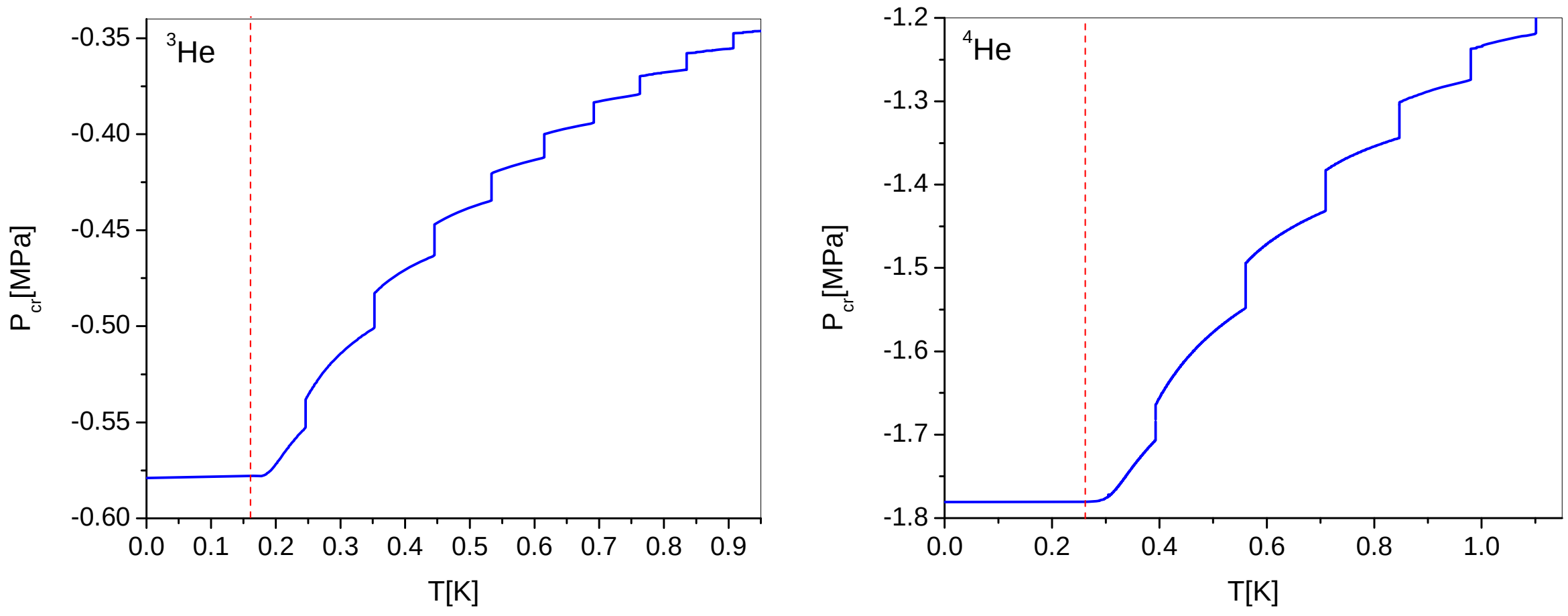

**Figure 5.** Calculated dependences of the critical pressure on temperature for $^3$He and $^4$He based on equation (11).

## Conclusions

In this paper, based on the model Schrödinger equation, we consider the tunneling mechanism of cavitation in liquid helium and obtain threshold values of negative pressure as a function of temperature for $^3$He and $^4$He. The results of calculating the surface tension coefficients for flat and curved interfaces obtained in the approximation of the Lenard-Jones interaction potential are presented. It is shown that the temperature dependence of the critical pressure (at which cavitation begins) is stepwise in nature. The calculations performed taking into account the curvilinear boundary and the stretching of the liquid in the field of negative pressure are in satisfactory agreement with the experimental data [1].

## Appendix 1. The surface tension coefficient estimate at a planar interface.

Classical potential 12-6 of Lennard-Jones [16] for nonpolar molecules has a view:

$$W = 4W_m\left(\left(\frac{r_0}{r}\right)^{12} - \left(\frac{r_0}{r}\right)^{6}\right). \tag{A1}$$

In the formula (A1) for helium in a gaseous phase, $W_m = 14.1 \cdot 10^{-23}$J and $r_0 = 0.256 \cdot 10^{-9}$m [17]. It was shown in [18, 19] that at a temperature of liquid He$^4$ close to zero, $r_0$ practically does not differ from the value in the gaseous phase, and the corresponding values for $W_m$ are $8.75 \cdot 10^{-23}$J [18] and $7.54 \cdot 10^{-23}$ J [19], which are close to the experimental value $W_m = 9.17 \cdot 10^{-23}$ J [20].

Liquid molecules in the bulk and on the surface interact not only with the nearest neighbors but also with markedly distant molecules that are located at distances significantly larger than the size of molecules and intermolecular gaps. Therefore, we will assume that, with respect to each selected molecule, water is a continuous medium with a fixed average density $\rho$. The effective radius and volume of the probe molecule's interaction with its nearest neighbors are

$$R_m = d_m, \tag{A2}$$

$$V_m = \frac{4}{3}\pi R_m^3 = \frac{4}{3}\pi d_m^3, \tag{A3}$$

and the density of the molecules are

$$n_W = \rho/m, \tag{A4}$$

where $m$ is the mass of the molecule.

In our consideration, the probe molecule is represented as a ball with diameter $d_m$ (A2) and volume $V_m$ (A3).

In this case, the total energy of intermolecular pair interactions for a probe molecule inside a liquid is

$$W_{in} = 4W_m n_w \int_{d_m=\alpha r_0}^{\infty} 4\pi\left(\left(\frac{r_0}{r}\right)^{12} - \left(\frac{r_0}{\mathrm{r}}\right)^{6}\right) r^2 dr = 4W_m n_w V_m \frac{1}{\alpha^6}\left(\frac{1}{3\alpha^6} - 1\right) = 4W_m n_w V_m I_0, \tag{A5}$$

where

$$I_0 = \frac{1}{\alpha^6}\left(\frac{1}{3\alpha^6} - 1\right). \tag{A6}$$

Now let us define the coefficient of the surface tension of the liquid. We define the coefficient of the surface tension of the liquid $\sigma$ as the difference between the potential energies of the probe molecule at the interface and inside the liquid (A5), integrated from the interface to infinite liquid depth and divided by the volume $V_m$ , Fig. A1.

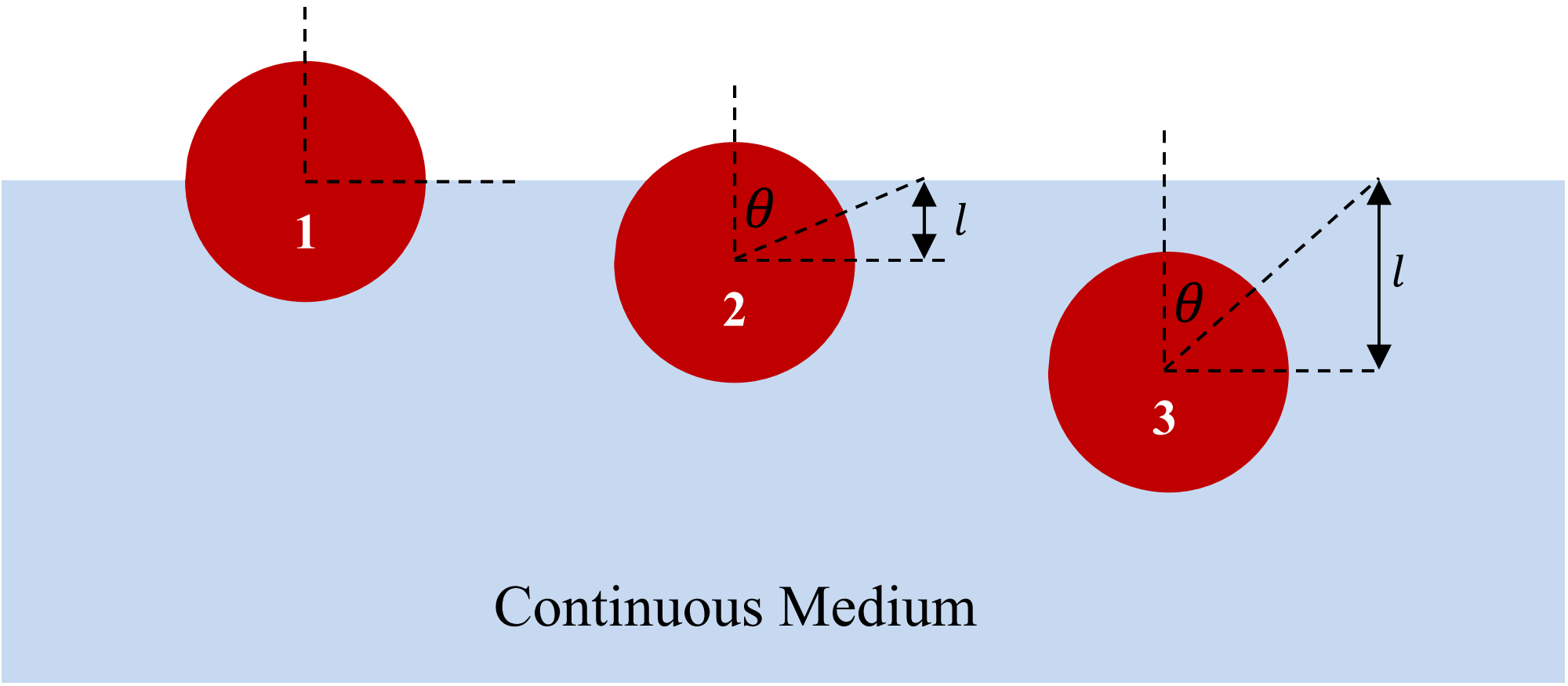

**Figure A1.** Illustration of the method for calculating the surface tension coefficient of a planar liquid–air interface. 1 – test molecule is on the liquid–air interface; 2 – test molecule is partially immersed in the liquid, $l \le d_m = \alpha r_0$; 3 – test molecule is completely immersed in the liquid, $l \ge d_m = \alpha r_0$.

In accordance with the definition of the surface tension coefficient (Fig. A1) is:

$$\sigma_0 = 4W_m n_w \left(\int_0^{d_m}\left(I_{l<d_m} - I_0\right)dl + \int_{d_m}^{\infty}\left(I_{l\ge d_m} - I_0\right)dl\right) = \frac{3}{2\alpha^6}\left(1 - \frac{1}{4\alpha^6}\right) W_m n_w d_m, \tag{A7}$$

where

$$I_{l<d_m}=\frac{2\pi}{V_m}\int_0^{l/d_m}dy\int_{d_m}^{l/y}\left(\left(\frac{r_0}{r}\right)^{12}-\left(\frac{r_0}{\mathrm{r}}\right)^{6}\right)r^2dr+\frac{I_0}{2}=$$

$$\frac{3}{2\alpha^3}\int_0^{l/d_m}dy\int_{\alpha}^{\alpha l/(d_m y)}(x^{-10}-x^{-6})dx+\frac{I_0}{2}=\frac{I_0}{2}-\frac{3}{8\alpha^6}\frac{l}{d_m}\left(1-\frac{2}{5\alpha^6}\right), \tag{A8}$$

$$I_{l\geq d_m}=\frac{2\pi}{V_m}\int_0^{1}dy\int_{d_m}^{l/y}\left(\left(\frac{r_0}{r}\right)^{12}-\left(\frac{r_0}{\mathrm{r}}\right)^{6}\right)r^2dr+\frac{I_0}{2}=$$

$$\frac{3}{2\alpha^3}\int_0^{1}dy\int_{\alpha}^{\alpha l/(d_m y)}(x^{-10}-x^{-6})dx+\frac{I_0}{2}=I_0+\left(\frac{1}{8\alpha^6}\frac{d_m^3}{l^3}-\frac{1}{60\alpha^{12}}\frac{d_m^9}{l^9}\right), \tag{A9}$$

where $y=\cos\theta$, and $x=\frac{r}{r_0}=\alpha\frac{r}{\mathrm{d}_m}$ (Fig. A1).

The index 0 in $\sigma_{0,LJ}$ indicates that the surface tension coefficient refers to the flat boundary, and the index $LJ$ is calculated on the basis of the Lennard-Jones model.

Considering that for liquid $^4$He the particle density is $n_W=2.18\cdot 10^{28}\ \mathrm{m}^{-3}$, substituting $W_m=9.17\cdot 10^{-23}\ \mathrm{J}$, $d_m=2^{1/6}\cdot r_0=0.287\cdot 10^{-9}\ \mathrm{m}$, $\alpha^6=2$, and $n_W$ in (A7), we obtain $\sigma_{0,s}=0.3764\cdot 10^{-3}\ \mathrm{N/m}$. This value practically coincides with the measured surface tension coefficient $\sigma_{0,s}=0.373\cdot 10^{-3}\mathrm{N/m}$ [21].

**Appendix 2. Surface tension of a bubble interface.**

Consider, to begin with, the gas–bubble interface in a liquid. Let us determine the surface tension coefficient at the boundary with a gas bubble, similar to the case of a planar interface. The energy difference between the probe molecule near the gas–liquid interface of the bubble and its energy inside the liquid is equal to the interaction energy of the probe molecule with the liquid, which would be occupied by the liquid in the bubble (Fig. A2).

In accordance with the definition of the surface tension, the coefficient for the bubble radius $R_b$ (Fig. A2) is

$$\sigma_b(R_b)=4W_m n_w\left(\int_{R_b}^{l_*}I_{b,1}dl+\int_{l_*}^{\infty}I_{b,2}dl\right), \tag{A10}$$

where:

$$l_*=\left(d_m^2+R_b^2\right)^{1/2}, \tag{A11}$$

$$I_{b,1}(l)=\frac{2\pi}{V_m}\int_{y_0}^{1}dy\int_{r_{1,b}}^{r_{1,e}}\left(\left(\frac{r_0}{r}\right)^{12}-\left(\frac{r_0}{\mathrm{r}}\right)^{6}\right)r^2dr=\frac{3}{2\alpha^3}\int_{y_0}^{1}dy\int_{r_{1,b}\alpha/d_m}^{r_{1,e}\,\alpha/d_m}\left(\frac{1}{x^{10}}-\frac{1}{x^4}\right)dx, \tag{A12}$$

$$I_{b,2}(l)=\frac{2\pi}{V_m}\int_{y_0}^{1}dy\int_{r_{2,b}}^{r_{2,e}}\left(\left(\frac{r_0}{r}\right)^{12}-\left(\frac{r_0}{\mathrm{r}}\right)^{6}\right)r^2dr=\frac{3}{2\alpha^3}\int_{y_0}^{1}dy\int_{r_{2,b}\alpha/d_m}^{r_{2,e}\,\alpha/d_m}\left(\frac{1}{x^{10}}-\frac{1}{x^4}\right)dx. \tag{A13}$$

In (A12), (A13) $y = \cos\theta$, and, correspondingly:

$$y_0 = \cos\theta_0 = \sqrt{1 - \frac{R_b^2}{l^2}} \tag{A14}$$

$$r_{b,1} = \max\left(d_m,\ l\cos\theta - \sqrt{R_b^2 - l^2\sin^2\theta}\right) \tag{A15}$$

$$r_{e,1} = l\cos\theta + \sqrt{R_b^2 - l^2\sin^2\theta} \tag{A16}$$

$$r_{b,2} = l\cos\theta - \sqrt{R_b^2 - l^2\sin^2\theta} \tag{A17}$$

$$r_{e,2} = l\cos\theta + \sqrt{R_b^2 - l^2\sin^2\theta}. \tag{A18}$$

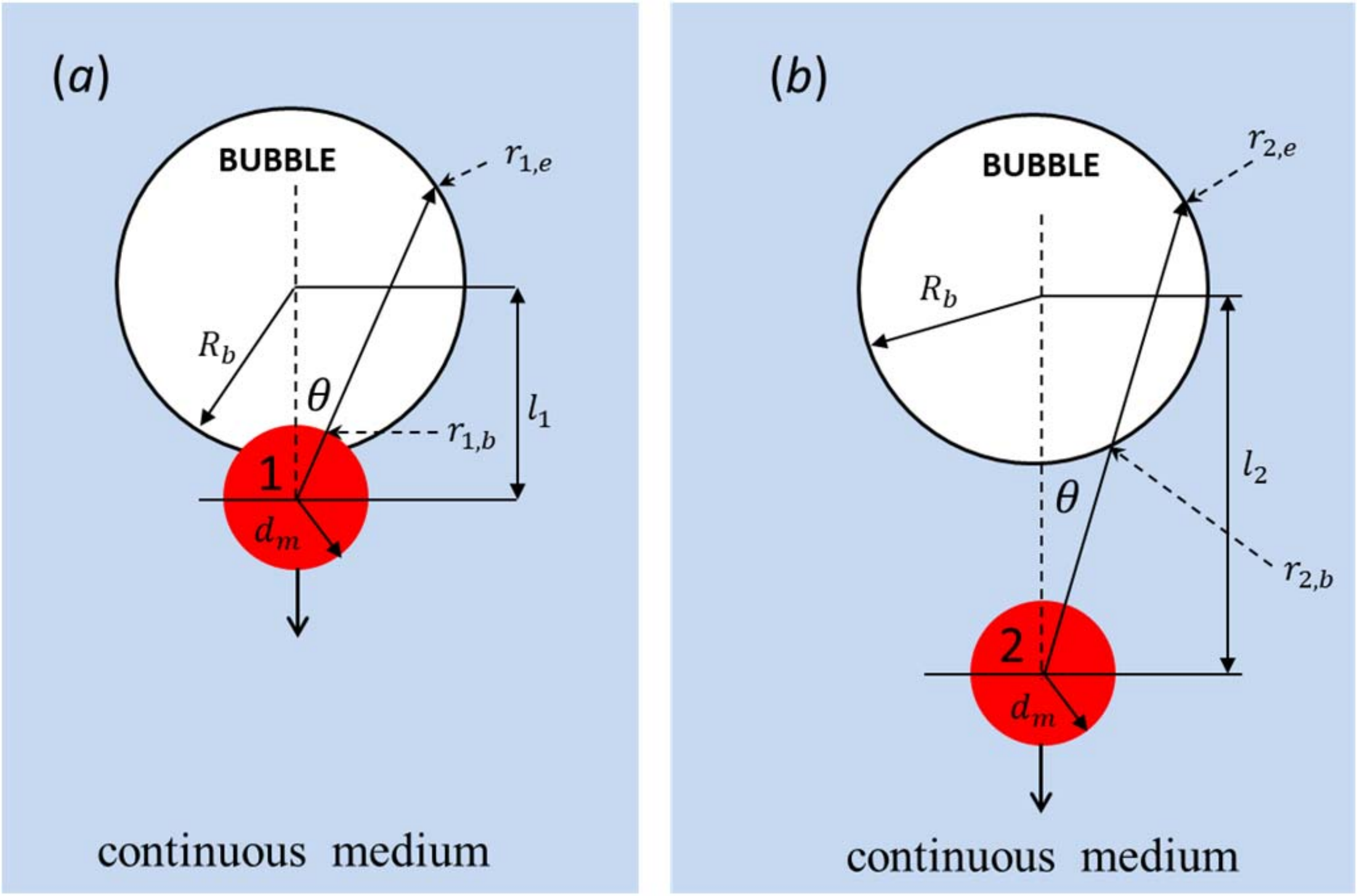

**Figure A2.** Illustration of the method of calculating the coefficient of the surface tension on the liquid–gas bubble interface. (a) Test molecule 1 is on the boundary of the bubble, and $l_1 < d_m + R_b$ is the distance from the center of the bubble to the center of the molecule. $r_{1,b}$, $r_{1,e}$ are the beginning and end of the line, respectively, along which integration in (A9) for molecule 1 occurs. (b) Test molecule 2 is completely immersed in the liquid, and $l_2 \geq d_m + R_b$, $r_{2,b}$, and $r_{2,e}$ are the beginning and end of the line, respectively, of integration for molecule 2.

The dependence of the surface tension coefficient $\sigma_b$ on the bubble radius $R_b$, calculated using formula (A10), is shown in Fig. A3. If the bubble radius increases, then $\sigma_b(R_b) \to \sigma_0$ at $R_b \to \infty$, where $\sigma_0, s$ is the surface tension coefficient for the planar interface.

Below we will approximate the surface tension coefficient of the bubble in liquid helium with curve 1 in Fig. A3, where $d_m = 0.287 \cdot 10^{-9}$ m, and $\sigma_0 = 3.73 \cdot 10^{-4}$ N/m and $1.5 \cdot 10^{-4}$ N/m for $^4$He [21] and $^3$He [22], respectively.

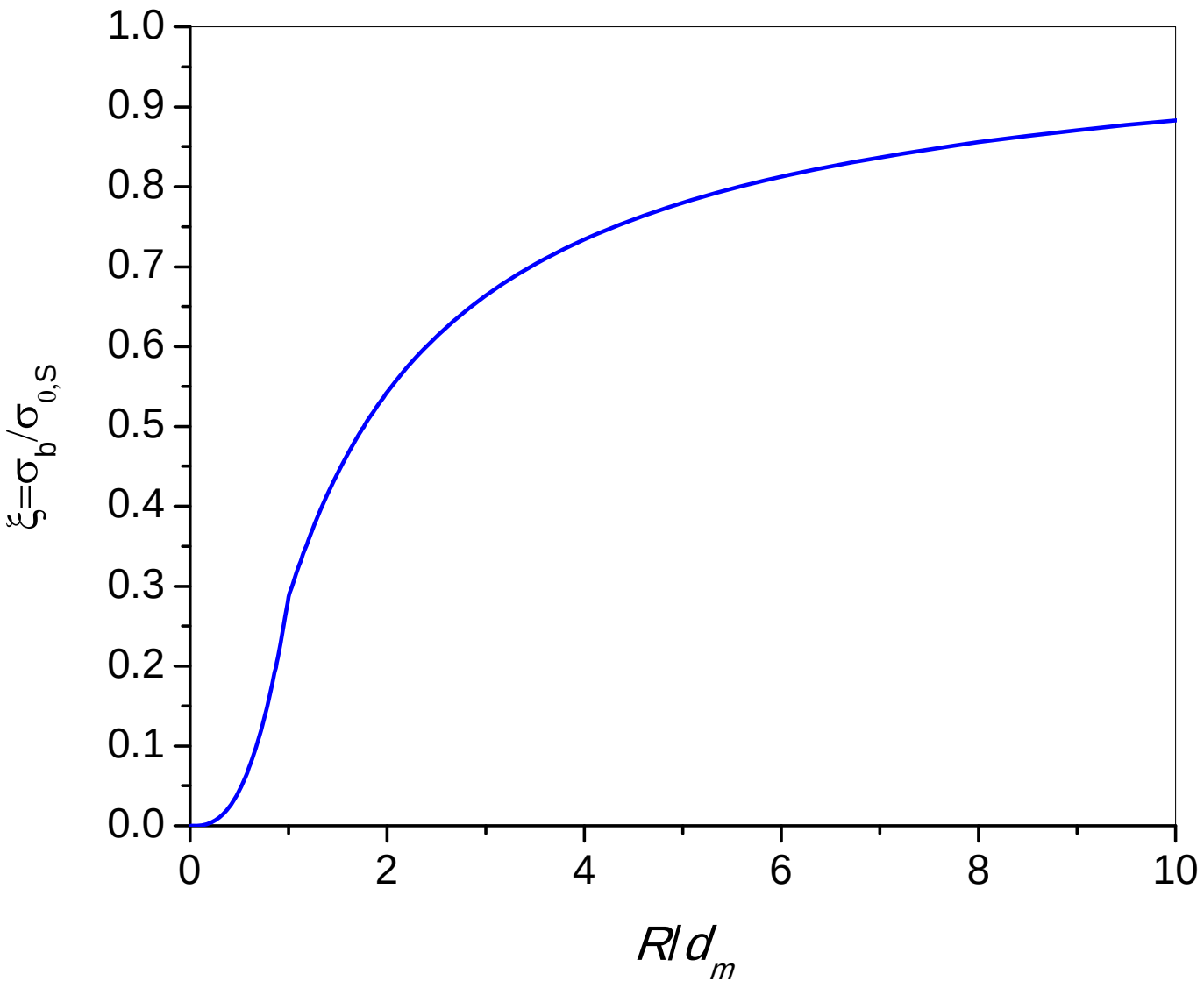

**Figure A3.** The dependences of the surface tension coefficient $\sigma_b$ normalized to the surface tension coefficient of a flat interface $\sigma_{0,s}$ on the bubble radius $R_b/d_m$.

**Appendix 3. Surface tension of a bubble in a negative pressure field.**

In the field of negative pressure, the liquid is stretched, and, accordingly, the value $n_w = \rho_{|P_-|}/m$ changes, where $\rho_{|P_-|}$ is the density of the liquid in the field of negative pressure $P_-$. The relation between the density of $^4$He and $^3$He and the negative pressure $P_-$ can be obtained by integrating the equations for the speed of sound $c^2$ [1]:

$$\left(\frac{\partial P}{\partial \rho}\right)_T = c^2_{He^4} = \left(14.3(P + 9.5 \cdot 10^5)\right)^{2/3} \quad (\mathrm{He}^4) \qquad \text{(A19)}$$

$$\left(\frac{\partial P}{\partial \rho}\right)_T = c^2_{He^3} = \left(19.23(P + 3 \cdot 10^5)\right)^{2/3} \quad (\mathrm{He}^3) \qquad \text{(A20)}$$

In (A19) and (A20) pressure $P$ is in Pa, and the sound speed of $c_{He^4}$ and $c_{He^3}$ are in m/s.

By counting the pressure from atmospheric pressure, we obtain the following equations of state for $^4$He and $^3$He; from (A19) and (A20) we have:

$$P_- = 7.57(\rho_{|P_-|} - 93.24)^3 - 9.5 \cdot 10^5 - p_{atm} = 7.57(\rho_{|P_-|} - 93.24)^3 - 10.5 \cdot 10^5 \quad {}^4\mathrm{He} \qquad \text{(A21)}$$

$$P_- = 13.7(\rho_{|P_-|} - 51.55)^3 - 3.0 \cdot 10^5 - p_{atm} = 13.7(\rho_{|P_-|} - 51.55)^3 - 4 \cdot 10^5 \quad {}^3\mathrm{He} \qquad \text{(A22)}$$

Using (A21) and (A22), we find the dependences of the density of liquid helium on the negative pressure and, accordingly, the surface tension coefficient for $\mathrm{He}^4$ and $\mathrm{He}^3$ in the case of a flat interface (Fig. A4).

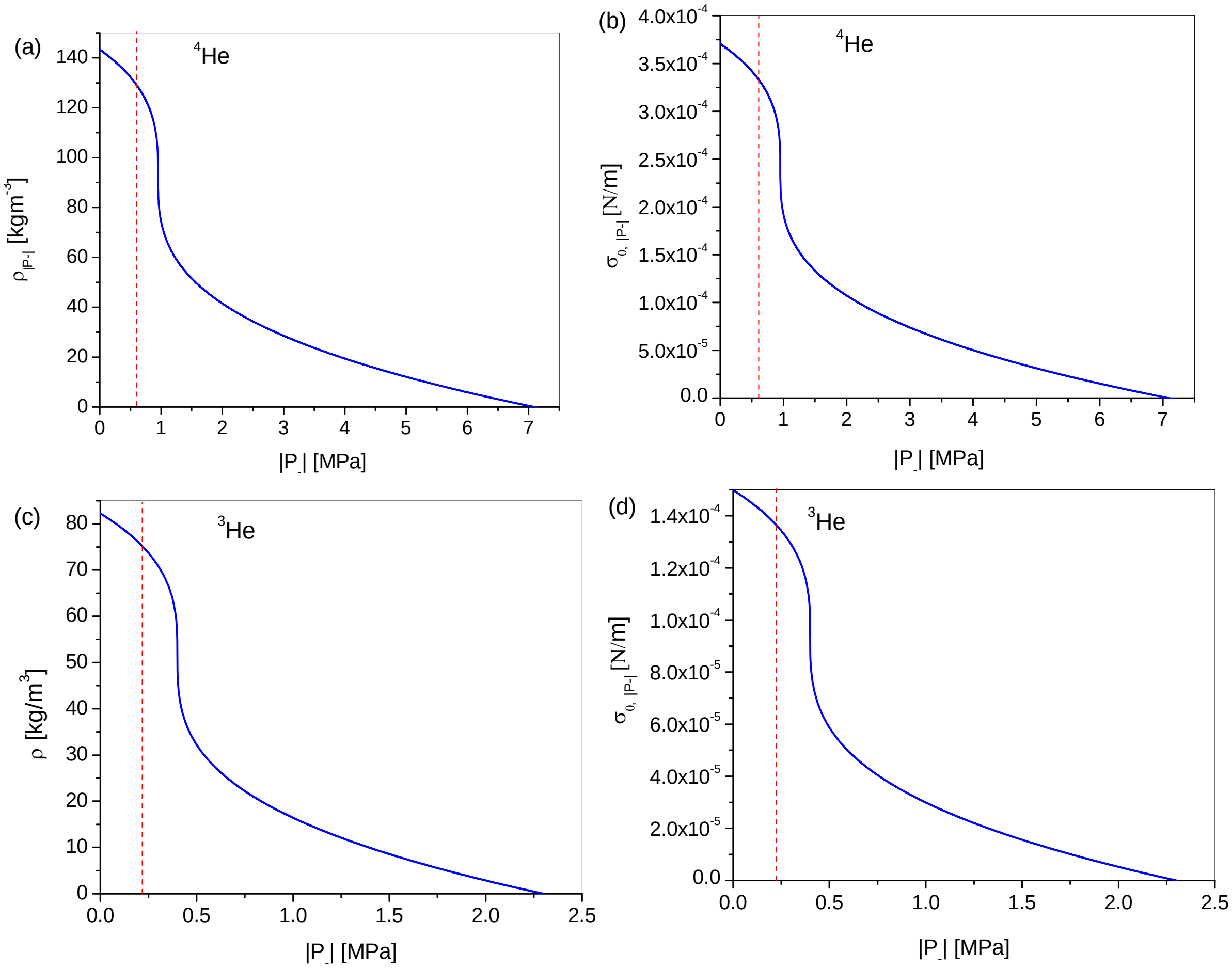

**Figure A4.** Dependences of density and coefficient of surface tension for a flat interface for liquid helium. (a) and (b) for $^4$He and (c) and (d) for $^3$He. Vertical red dashed lines show the maximum absolute values of negative pressure at which cavitation begins (see Fig. 4).